\documentclass[journal]{IEEEtran}
\usepackage{amsmath}
\usepackage{graphicx}
\usepackage{amsfonts}
\usepackage{amssymb}
\usepackage{alltt}
\usepackage{booktabs}
\usepackage{stmaryrd}
\usepackage{dsfont} 
\usepackage{color} 
\usepackage[standard]{ntheorem}
\usepackage[font=footnotesize]{subfig}
\usepackage{array,algorithm2e,algorithmic}
\renewcommand{\theequation}{\arabic{equation}}
\usepackage[width=18.2cm,height=25.7cm]{geometry}
\usepackage{psfig}
\usepackage{epsfig}
\ifCLASSINFOpdf
\else
\fi

\begin{document}

\title{Evaluating Quality of Chaotic Pseudo-Random Generators: Application to Information Hiding}
\author{\IEEEauthorblockN{Jacques M. Bahi,
Xiaole Fang,
Christophe Guyeux, and
Qianxue Wang\\}
\IEEEauthorblockA{University of Franche-Comt\'e\\
Computer Science Laboratory LIFC,
Besan\c con, France\\ Email:\{jacques.bahi, xiaole.fang,
christophe.guyeux, qianxue.wang\}@univ-fcomte.fr }}
\maketitle
\thispagestyle{empty}

\begin{abstract}
Guaranteeing the security of information transmitted through the Internet, against passive or active attacks, is a major concern. 
The discovery of new pseudo-random number generators with a strong level of security is a field of research in full expansion, due to the fact that numerous cryptosystems and data hiding schemes are directly dependent on the quality of these generators.
At the conference Internet`09, we described a generator based on chaotic iterations which behaves chaotically as defined by Devaney. In this paper which is an extension of the work presented at the conference Internet`10, the proposal is to improve the speed, the security, and the evaluation of this generator, to make its use more relevant in the Internet security context. In order to do so, a comparative study between various generators is carried out and statistical results are improved. Finally, an application in the information hiding framework is presented with details, to give an illustrative example of the use of such a generator in the Internet security field. 
\end{abstract}
\IEEEpeerreviewmaketitle

\section{Introduction}

Due to the rapid development of the Internet in recent years, the need to find new tools to reinforce trust and security through the Internet has become a major concern. 
Its recent role in everyday life implies the need to protect data and privacy in digital world. This extremely rapid development of the Internet brings more and more attention to the information security techniques in all kinds of applications. For example, new security concerns have recently appeared because of the evolution of the Internet to support such activities as e-Voting, VoD (Video on demand), and the protection of intellectual property. 
In all these emerging techniques, pseudo-random number generators (PRNG) play an important role, because they are fundamental components of almost all cryptosystems and information hiding schemes~\cite{Tong2009480,Erclebi2005}.
PRNGs are typically defined by a deterministic recurrent sequence in a finite state space, usually a finite field or ring, and an output function mapping each state to an input value. Following~\cite{L'ecuyer2008}, this value is often either a real number in the interval $(0,1)$ or an integer in some finite range. PRNGs based on linear congruential methods and feedback shift-registers are popular for historical reasons~\cite{Knuth1998}, but their security level often has been revealed to be inadequate by today's standards.
However, to use a PRNG with a high level of security is a necessity to protect the information contents sent through the Internet.
This level depends both on theoretical properties and on statistical tests.

Many PRNGs have already been proven to be secure following a probabilistic approach~\cite{Marchi20093328,Sachez2005,Tan2002693}. However, their performances must regularly be improved, among other things by using new mathematical tools.
This is why the idea of using chaotic dynamical systems for this purpose has recently been explored~\cite{Falcioni2005,Cecen2009}.
The random-like and unpredictable dynamics of chaotic systems, their inherent determinism and simplicity of realization suggest their potential for exploitation as PRNGs.
Such generators can strongly improve the confidence put in any information hiding scheme and in cryptography in general: due to their properties of unpredictability, the possibilities offered to an attacker to achieve his goal are drastically reduced in that context.
For example, in cryptography, keys are needed to be unpredictable enough, to make sure any search optimization based on the reduction of the key space to the most probable values is impossible to work on.
But the number of generators claimed as chaotic, which actually have been proven to be unpredictable (as it is defined in the mathematical theory of chaos) is very small.

\section{Outline of our Work}

This paper extends the study initiated in~\cite{A2010,guyeux09,wang2009}, and tries to fill this gap. In~\cite{guyeux09}, it is mathematically proven that chaotic iterations (CIs), a suitable tool for fast computing distributed algorithms, satisfies the topological chaotic property, following the definition given by Devaney~\cite{Dev89}.
In the paper~\cite{wang2009} presented at Internet`09, the chaotic behavior of CIs is exploited in order to obtain an unpredictable PRNG that depends on two logistic maps.
We have shown that, in addition to being chaotic, this generator can pass the NIST (National Institute of Standards and Technology of the U.S. Government) battery of tests~\cite{ANDREW2008},
widely considered as a comprehensive and stringent battery of tests for cryptographic applications.
In this paper, which is an extension of \cite{A2010}, we have improved the speed, security, and evaluation of the former generator and of its application in information hiding.
Chaotic properties, statistical tests, and security analysis~\cite{ZHENG92008} allow us to consider that this generator has good characteristics and is capable to withstand attacks.
After having presented the theoretical framework of the study and a security analysis, we will give a comparison based on statistical tests. Finally a concrete example of how to use these pseudo-random numbers for information hiding through the Internet is detailed.

The remainder of this paper is organized in the following way. In Section~\ref{Basic recalls}, some basic definitions concerning chaotic iterations and PRNGs are recalled. Then, the generator based on discrete chaotic iterations is presented in Section~\ref{The generation of pseudo-random sequence}. Section~\ref{Security analysis} is devoted to its security analysis. In Section~\ref{Comparative analysis}, various tests are passed with a goal to achieve a statistical comparison between this new PRNG and other existing ones. In Section~\ref{An application example of the proposed PRNG}, a potential use of this PRNG in some Internet security field is presented, namely in information hiding. The paper ends with a conclusion and intended future work.

\section{Review of Basics}
\label{Basic recalls}

\subsection{Notations}
\begin{tabular}{@{}c@{}@{}l@{}}
$\llbracket 1;\mathsf{N} \rrbracket$ & $\rightarrow\{1,2,\hdots,\mathsf{N}\}$ \\
$S^{n}$ & $\rightarrow$ the $n^{th}$ term of a sequence $S=(S^{1},S^{2},\hdots)$ \\
$v_{i}$ & $\rightarrow$ the $i^{th}$ component of a vector \\
&~~~~$v=(v_{1},v_{2},\hdots, v_n)$\\
$f^{k}$ & $\rightarrow$ $k^{th}$ composition of a function $f$ \\
$\emph{strategy}$~ & $\rightarrow$ a sequence which elements belong in $%
\llbracket 1;\mathsf{N} \rrbracket $ \\
$\mathbb{S}$ & $\rightarrow$ the set of all strategies \\
$\mathbf{C}_n^k$ & $\rightarrow$ the binomial coefficient ${n \choose k} = \frac{n!}{k!(n-k)!}$\\
$\oplus$ & $\rightarrow$ bitwise exclusive or \\
$+$ & $\rightarrow$ the integer addition \\
$\ll \text{and} \gg$ & $\rightarrow$ the usual shift operators \\
$(\mathcal{X}, \text{d})$ & $\rightarrow$ a metric space \\
$mod$ & $\rightarrow$ a modulo or remainder operator
\end{tabular}

\begin{tabular}{@{}c@{}@{}l@{}}

$\lfloor x \rfloor$ & $\rightarrow$ returns the highest integer smaller than $x$ \\
$n!$ & $\rightarrow$ the factorial $n!=n\times(n-1)\times\dots\times1$\\
$\mathds{N}^{\ast }$ & $\rightarrow$ the set of positive integers \{1,2,3,...\}
\end{tabular}

\subsection{XORshift}

XORshift is a category of very fast PRNGs designed by George Marsaglia~\cite{Marsaglia2003}.
It repeatedly uses the transform of exclusive or (XOR) on a number with a bit shifted version of it. The state of a XORshift generator is a vector of bits. At each step, the next state is obtained by applying a given number of XORshift operations to $w$-bit blocks in the current state, where $w = 32$ or $64$. A XORshift operation is defined as follows. Replace the $w$-bit block by a bitwise XOR of the original block, with a shifted copy of itself by $a$ positions either to the right or to the left, where $ 0 < a < w$. This Algorithm~\ref{XORshift} has a period of $2^{32}-1=4.29\times10^9$.

\begin{algorithm}
\SetAlgoLined
\KwIn{the internal state $z$ (a 32-bit word)}
\KwOut{$y$ (a 32-bit word)}
$z\leftarrow{z\oplus{(z\ll13)}}$\;
$z\leftarrow{z\oplus{(z\gg17)}}$\;
$z\leftarrow{z\oplus{(z\ll5)}}$\;
$y\leftarrow{z}$\;
return $y$\;
\medskip
\caption{An arbitrary round of XORshift algorithm}
\label{XORshift}
\end{algorithm}

\subsection{Continuous Chaos in Digital Computers}

In the past two decades, the use of chaotic systems in the design of cryptosystems, pseudo-random number generators (PRNG), and hash functions, has become more and more frequent.
Generally speaking, the chaos theory in the continuous field is used to analyze performances of related systems. However, when chaotic systems are realized in digital computers with finite computing precisions, it is doubtful whether or not they can still preserve the desired dynamics of the continuous chaotic systems. Because most dynamical properties of chaos are meaningful only when dynamical systems evolve in the continuous phase space, these properties may become meaningless or ambiguous when the phase space is highly quantized (i.e., latticed) with a finite computing precision (in other words, dynamical degradation of continuous chaotic systems realized
in finite computing precision). When chaotic systems are realized in finite precision, their dynamical properties will be deeply different from the properties of continuous-value systems and some dynamical degradation will arise, such as short cycle length and decayed distribution. This phenomenon has been reported and analyzed in various situations~\cite{Binder1986,Wheeler1989,Palmore1990,Blank1997,Li2005}.

Therefore, continuous chaos may collapse into the digital world and the ideal way to generate pseudo-random sequences is to use a discrete-time chaotic system.

\subsection{Chaos for Discrete Dynamical Systems}

Consider a metric space $(\mathcal{X},d)$ and a continuous function $f:\mathcal{X}\longrightarrow \mathcal{X}$, for one-dimensional dynamical systems of the form:
\begin{equation}
x^0 \in \mathcal{X} \textrm{  and } \forall n \in \mathds{N}^*, x^n=f(x^{n-1}),
\label{Devaney}
\end{equation}
the following definition of chaotic behavior, formulated by Devaney~\cite{Dev89}, is widely accepted:

\begin{definition}
 A dynamical system of Form~(\ref{Devaney}) is said to be chaotic if the following conditions hold.
\begin{itemize}
\item Topological transitivity:
\begin{equation}
\forall U,V \textrm{ open sets of } \mathcal{X}\setminus \varnothing, ~\exists k>0, f^k(U) \cap V \neq \varnothing
\end{equation}
\item Density of periodic points in $\mathcal{X}$:

Let $P=\{p\in \mathcal{X}|\exists n \in \mathds{N}^{\ast}:f^n(p)=p\}$ the set of periodic points of $f$. Then $P$ is dense in $\mathcal{X}$:
\begin{equation}
 \overline{P}=\mathcal{X}
\end{equation}

\item Sensitive dependence on initial conditions:
$\exists \varepsilon>0,$ $\forall x \in \mathcal{X},$ $\forall \delta >0,$ $\exists y \in \mathcal{X},$ $\exists n \in \mathbb{N},$ $d(x,y)<\delta$ and $d\left(f^n(x),f^n(y)\right) \geqslant \varepsilon.$
\end{itemize}

\end{definition}
When $f$ is chaotic, then the system $(\mathcal{X}, f)$ is chaotic and quoting Devaney: ``it is unpredictable because of the sensitive dependence on initial conditions. It cannot be broken down or decomposed into two subsystems which do not interact because of topological transitivity. And, in the midst of this random behavior, we nevertheless have an element of regularity.'' Fundamentally different  behaviors  are  consequently possible and occur in an unpredictable way.

\subsection{Discrete Chaotic Iterations}
\label{Chaotic iterations}

\begin{definition}
The set $\mathds{B}$ denoting $\{0,1\}$, let $f:\mathds{B}^{\mathsf{N}%
}\longrightarrow \mathds{B}^{\mathsf{N}}$ be an ``iteration'' function and $S\in \mathbb{S}
$ be a chaotic strategy. Then, the so-called \emph{chaotic iterations}~\cite{Robert1986} are defined by $x^0\in \mathds{B}^{\mathsf{N}}$ and
\begin{equation}
\forall n\in \mathds{N}^{\ast },\forall i\in \llbracket1;\mathsf{N}\rrbracket%
,x_i^n=\left\{
\begin{array}{l}
x_i^{n-1}~~~~~\text{if}~S^n\neq i \\
f(x^{n-1})_{S^n}~\text{if}~S^n=i.\end{array} \right. 
\end{equation}
\end{definition}
In other words, at the $n^{th}$ iteration, only the $S^{n}-$th cell is
\textquotedblleft iterated\textquotedblright . Note that in a more general
formulation, $S^n$ can be a subset of components and $f(x^{n-1})_{S^{n}}$ can
be replaced by $f(x^{k})_{S^{n}}$, where $k < n$, describing for
example delays transmission. For the
general definition of such chaotic iterations, see, e.g.,~\cite{Robert1986}.

Chaotic iterations generate a set of vectors (Boolean vector in this paper),
they are defined by an initial state $x^{0}$, an iteration function $f$, and a chaotic strategy $S$.

The next section gives the outline proof that chaotic iterations satisfy Devaney's topological chaos property. Thus they can be used to define a chaotic pseudo-random bit generator.

%
%
%
%
%
%
%

\section{The Generation of CI Pseudo-Random Sequence}
\label{The generation of pseudo-random sequence}

\subsection{A Theoretical Proof for Devaney's Chaotic Dynamical Systems}
\label{A theoretical proof for Devaney's chaotic dynamical systems}
The outline proofs, of the properties on which our pseudo-random number generator is based, are given in this section.

Denote by $\delta $ the \emph{discrete Boolean metric}, $\delta
(x,y)=0\Leftrightarrow x=y.$ Given a function $f$, define the function $%
F_{f}:$ $\llbracket1;\mathsf{N}\rrbracket\times \mathds{B}^{\mathsf{N}%
}\longrightarrow \mathds{B}^{\mathsf{N}}$ such that $$F_{f}(k,E)=\left(
E_{j}.\delta (k,j)+f(E)_{k}.\overline{\delta (k,j)}\right) _{j\in \llbracket%
1;\mathsf{N}\rrbracket},$$ where + and . are the Boolean addition and product operations.

Consider the phase space: $\mathcal{X}=\llbracket1;\mathsf{N}\rrbracket^{%
\mathds{N}}\times \mathds{B}^{\mathsf{N}}$ and the map $$G_{f}\left( S,E\right) =\left( \sigma
(S),F_{f}(i(S),E)\right) ,$$ then the chaotic iterations defined in (\ref{Chaotic iterations}) can be described by the following iterations \cite{guyeux09}
\[
\left\{
\begin{array}{l}
X^{0}\in \mathcal{X} \\
X^{k+1}=G_{f}(X^{k}).%
\end{array}%
\right.
\]

Let us define a new distance between two points $(S,E),(\check{S},\check{E})\in
\mathcal{X}$ by $$d((S,E);(\check{S},\check{E}))=d_{e}(E,\check{E})+d_{s}(S,%
\check{S}),$$ where
\begin{itemize}
\item $\displaystyle{d_{e}(E,\check{E})}=\displaystyle{%
\sum_{k=1}^{\mathsf{N}}\delta (E_{k},\check{E}_{k})} \in \llbracket 0 ; \mathsf{N} \rrbracket$ \\
\item $\displaystyle{%
d_{s}(S,\check{S})}=\displaystyle{\dfrac{9}{\mathsf{N}}\sum_{k=1}^{\infty }%
\dfrac{|S^{k}-\check{S}^{k}|}{10^{k}}} \in [0 ; 1].$
\end{itemize}

\medskip

It is then proven in \cite{guyeux09} by using the sequential continuity that

\begin{proposition}
\label{continuite} $G_f$ is a continuous function on $(\mathcal{X},d)$.
\end{proposition}

Then, the vectorial negation $f_{0}(x_{1},%
\hdots,x_{\mathsf{N}})=(\overline{x_{1}},\hdots,\overline{x_{\mathsf{N}}})$ satisfies the three conditions for Devaney's chaos, namely, regularity, transitivity, and sensitivity in the metric space $(\mathcal{X},d)$. This leads to the following result.

\begin{proposition}
$G_{f_0}$ is a chaotic map on $(\mathcal{X},d)$ in the sense of Devaney.
\end{proposition}

\subsection{Chaotic Iterations as Pseudo-Random Generator}
\subsubsection{Presentation}
The CI generator (generator based on chaotic iterations) is designed by the following process. First of all, some chaotic iterations have to be done to generate a sequence $\left(x^n\right)_{n\in\mathds{N}} \in \left(\mathds{B}^\mathsf{N}\right)^\mathds{N}$ ($\mathsf{N} \in \mathds{N}^*, \mathsf{N} \geqslant 2$, $\mathsf{N}$ is not necessarily equal to 32) of Boolean vectors, which are the successive states of the iterated system. Some of these vectors will be randomly extracted and our pseudo-random bit flow will be constituted by their components. Such chaotic iterations are realized as follows. Initial state $x^0 \in \mathds{B}^\mathsf{N}$ is a Boolean vector taken as a seed (see Section~\ref{algo seed}) and chaotic strategy $\left(S^n\right)_{n\in\mathds{N}}\in \llbracket 1, \mathsf{N} \rrbracket^\mathds{N}$ is
an irregular decimation of a XORshift sequence (Section~\ref{Chaotic strategy}). The iterate function $f$ is
the vectorial Boolean negation:
$$f_0:(x_1,...,x_\mathsf{N}) \in \mathds{B}^\mathsf{N} \longmapsto (\overline{x_1},...,\overline{x_\mathsf{N}}) \in \mathds{B}^\mathsf{N}.$$
At each iteration, only the $S^i$-th component of state $x^n$ is updated, as follows: $x_i^n = x_i^{n-1}$ if $i \neq S^i$, else $x_i^n = \overline{x_i^{n-1}}$.
Finally, some $x^n$ are selected
by a sequence $m^n$ as the pseudo-random bit sequence of our generator.
$(m^n)_{n \in \mathds{N}} \in \mathcal{M}^\mathds{N}$ is computed from a XORshift sequence $(y^n)_{n \in \mathds{N}} \in \llbracket 0, 2^{32}-1 \rrbracket$ (see Section~\ref{algo m}). So, the
generator returns the following values:\newline
\begin{small}
Bits:$$x_1^{m_0}x_2^{m_0}x_3^{m_0}\hdots x_\mathsf{N}^{m_0}x_1^{m_0+m_1}x_2^{m_0+m_1}\hdots x_\mathsf{N}^{m_0+m_1} x_1^{m_0+m_1+m_2}\hdots$$
or States:$$x^{m_0}x^{m_0+m_1}x^{m_0+m_1+m_2}\hdots$$
\end{small}

\subsubsection{The seed}
\label{algo seed}
The unpredictability of random sequences is established using
a random seed that is obtained by a physical source like timings of keystrokes.
Without the seed, the attacker must not be able to make any predictions about
the output bits, even when all details of the generator are known~\cite{Turan2008}.

The initial state of the system $x^0$ and the first term $y^0$ of the XORshift are seeded either by
the current time in seconds since the Epoch, or by a number that the user inputs.
Different ways are possible. For example, let us denote by $t$ the decimal part of the current
time. So $x^0$ can be $t \text{ (mod $2^N$)}$ written in binary digits and $y^0 = t$.

\subsubsection{Sequence $m$ of returned states}
\label{algo m}
The output of the sequence $(y^n)$ is uniform in $\llbracket 0, 2^{32}-1 \rrbracket$, because it is produced by a XORshift generator. However, we do not want the output of $(m^n)$ to be uniform in $\llbracket 0, N \rrbracket$, because in this case, the returns of our generator will not be uniform in $\llbracket 0, 2^{\mathsf{N}}-1 \rrbracket$, as it is illustrated in the following example. Let us suppose that $x^0=(0,0,0)$. Then $m^0 \in \llbracket 0, 3 \rrbracket$.
\begin{itemize}
\item If $m^0=0$, then no bit will change between the first and the second output of our PRNG. Thus $x^1 = (0,0,0)$.
\item If $m^0=1$, then exactly one bit will change, which leads to three possible values for $x^1$, namely $(1,0,0)$, $(0,1,0)$, and $(0,0,1)$.
\item \emph{etc.}
\end{itemize}
As each value in $\llbracket 0, 2^3-1 \rrbracket$ must be returned with the same probability, then the values $(0,0,0)$, $(1,0,0)$, $(0,1,0)$, and $(0,0,1)$ must occur for $x^1$ with the same probability. Finally we see that, in this example, $m^0=1$ must be three times more probable than $m^0=0$.
This leads to the following general definition for $m$:
\begin{equation}
\label{Formula}
m^n = g_1(y^n)=
\left\{
\begin{array}{l}
0 \text{ if }0 \leqslant\frac{y^n}{2^{32}}<\frac{C^0_N}{2^\mathsf{N}},\\
1 \text{ if }\frac{C^0_\mathsf{N}}{2^\mathsf{N}} \leqslant\frac{y^n}{2^{32}}<\sum_{i=0}^1\frac{C^i_N}{2^\mathsf{N}},\\
2 \text{ if }\sum_{i=0}^1\frac{C^i_\mathsf{N}}{2^\mathsf{N}} \leqslant\frac{y^n}{2^{32}}<\sum_{i=0}^2\frac{C^i_\mathsf{N}}{2^\mathsf{N}},\\
\vdots~~~~~ ~~\vdots~~~ ~~~~\\
\mathsf{N} \text{ if }\sum_{i=0}^{\mathsf{N}-1}\frac{C^i_\mathsf{N}}{2^\mathsf{N}} \leqslant\frac{y^n}{2^{32}}<1.\\
\end{array}
\right.
\end{equation}
%
%
%
%
%
%
%

\begin{figure}[!t]
\centering
\subfloat [$m^n = f(y^n)$]{\includegraphics[scale=0.4]{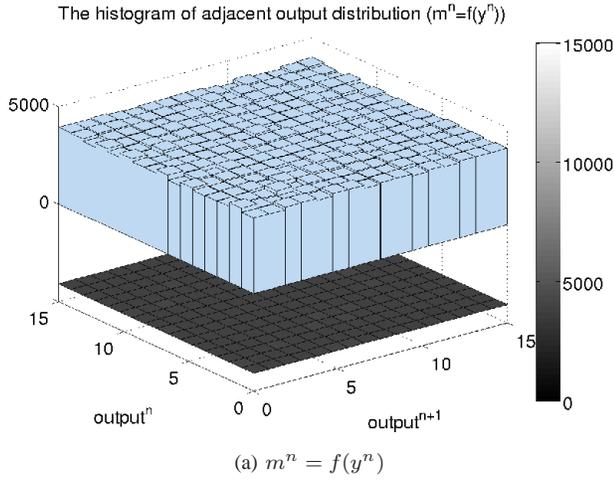}}

\subfloat [$m^n = y^n ~ mod ~ 4$]{\includegraphics[scale=0.4]{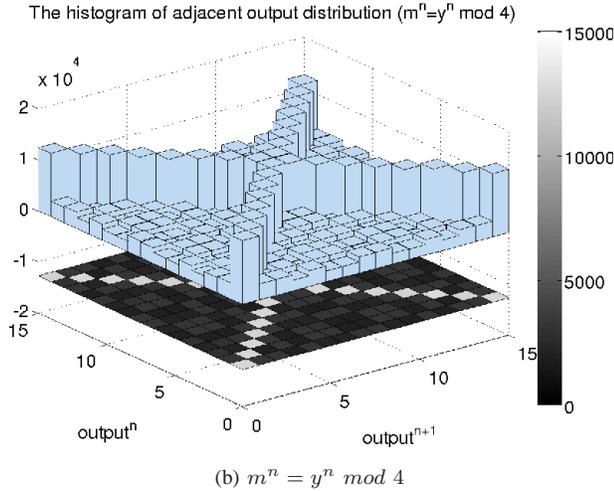}%
}
\caption{Histogram and intensity maps}
\label{Histogram and intensity map}
\end{figure}

In order to evaluate our proposed method and compare its statistical properties with various other methods, the density histogram and intensity map of adjacent outputs have been computed. The length of $x$ is $\mathsf{N} = 4$ bits, and the initial conditions and control
parameters are the same. A large number of
sampled values are simulated ($10^6$ samples). 
Figure~\ref{Histogram and intensity map}(a) shows the intensity map for $m^n=g_1(y^n)$.
In order to appear random, the histogram should be uniformly distributed in all areas. 
It can be observed that a uniform histogram and a flat color intensity map are obtained when using our scheme. 
Another illustration of this fact is given by Figure~\ref{Histogram and intensity map}(b), whereas its uniformity is further justified by the tests presented in Section \ref{Comparative analysis}.

\subsubsection{Chaotic strategy}
\label{Chaotic strategy}
The chaotic strategy $(S^k) \in \llbracket 1, \mathsf{N} \rrbracket^\mathds{N}$ is generated from a second XORshift sequence $(b^k) \in \llbracket 1, N \rrbracket^\mathds{N}$. The only difference between the sequences $S$ and $b$ is that some terms of $b$ are discarded, in such a way that $\forall k \in \mathds{N}, (S^{M^k}, S^{M^k+1}, \hdots, S^{M^{k+1}-1})$ does not contain any given integer twice, where $M^k = \sum_{i=0}^k m^i$. Therefore, no bit will change more than once between two successive outputs of our PRNG, increasing the speed of the former generator by doing so. $S$ is said to be ``an irregular decimation'' of $b$. This decimation can be obtained by the following process.

Let $(d^1,d^2,\dots,d^\mathsf{N})\in \{0,1\}^\mathsf{N}$ be a mark sequence, such that whenever $\sum_{i=1}^\mathsf{N} d^i = m^k$,
then $\forall i, d_i=0$ ($\forall k$, the sequence is reset when $d$ contains $m^k$ times the number 1). This mark sequence will control the XORshift sequence $b$ as follows:
\begin{itemize}
\item if $d^{b^j} \neq 1$, then $S^k=b^j$, $d^{b^j} = 1$, and $k = k+1$,
\item if $d^{b^j}=1$, then $b^j$ is discarded.
\end{itemize}
For example, if $b = 142\underline{2}334 1421\underline{1}\underline{2}\underline{2}34...$ and $m = 4341...$, then $S=1423~341~4123~4...$ However, if we do not use the mark sequence, then one position may change more than once and the balance property will not be checked, due to the fact that $\bar{\bar{x}}=x$. As an example, for $b$ and $m$ as in the previous example, $S=1422~334~1421~1...$ and $S=14~4~42~1...$ lead to the same outputs (because switching the same bit twice leads to the same state).

\begin{figure}
\centering
\includegraphics[width=3.85in]{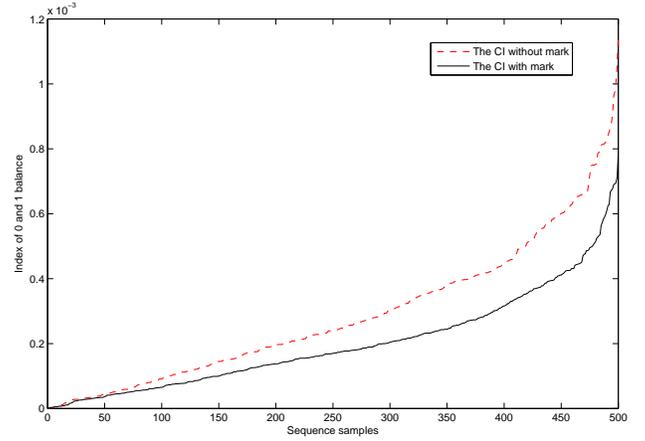}
\DeclareGraphicsExtensions.
\caption{Balance property}
\label{nmark}
\end{figure}

To check the balance property, a set of 500
sequences are generated with and without decimation, each
sequence containing $10^6$ bits. Figure~\ref{nmark} shows the
percentages of differences between zeros and ones, and presents a better balance property for the sequences with decimation. This claim will be verified in the tests section (Section \ref{Comparative analysis}).

Another example is given in Table~\ref{table application example}, in which $r$ means ``reset'' and the integers which are underlined in sequence $b$ are discarded.
%

\subsection{CI(XORshift, XORshift) Algorithm}

The basic design procedure of the novel generator is summed up in Algorithm~\ref{Chaotic iteration1}.
The internal state is $x$, the output state is $r$. $a$ and $b$ are those computed by the two XORshift
generators. The value $g_1(a)$ is an integer, defined as in Equation~\ref{Formula}. Lastly, $\mathsf{N}$ is a constant defined by the user.
\begin{algorithm}
\SetAlgoLined
\KwIn{the internal state $x$ ($\mathsf{N}$ bits)}
\KwOut{a state $r$ of $\mathsf{N}$ bits}
\For{$i=0,\dots,\mathsf{N}$}
{
$d_i\leftarrow{0}$\;
}
$a\leftarrow{XORshift1()}$\;
$m\leftarrow{g_1(a)}$\;
$k\leftarrow{m}$\;
\For{$i=0,\dots,k$}
{
$b\leftarrow{XORshift2()~mod~\mathsf{N}}$\;
$S\leftarrow{b}$\;
\If{$d_S=0$}
{
$x_S\leftarrow{ \overline{x_S}}$\;
$d_S\leftarrow{1}$\;
}
\ElseIf{$d_S=1$}
{
$k\leftarrow{ k+1}$\;
}
}
$r\leftarrow{x}$\;
return $r$\;
\medskip
\caption{An arbitrary round of the new CI(XORshift,XORshift) generator}
\label{Chaotic iteration1}
\end{algorithm}

As a comparison, the basic design procedure of the old generator is recalled in Algorithm~\ref{Chaotic iteration2} ($a$ and $b$ are computed by logistic maps, $\mathsf{N}$ and $c\geqslant 3\mathsf{N}$ are constants defined by the user). See~\cite{wang2009} for further information.

\begin{algorithm}
\SetAlgoLined
\KwIn{the internal state $x$ ($\mathsf{N}$ bits)}
\KwOut{a state $r$ of $\mathsf{N}$ bits}
$a\leftarrow{Logistic map1()}$\;
\If{$a>0.5$}
{
$d\leftarrow 1$
}
\Else
{
$d\leftarrow 0$
}

$m\leftarrow{d+c}$\;
\For{$i=0,\dots,m$}
{
$b\leftarrow{Logistic map2()}$\;
$S\leftarrow{100000b~mod~\mathsf{N}}$\;
$x_S\leftarrow{ \overline{x_S}}$\;
}
$r\leftarrow{x}$\;
return $r$\;
\medskip
\caption{An arbitrary round of the old CI PRNG}
\label{Chaotic iteration2}
\end{algorithm}

\subsection{Illustrative Example}

In this example, $\mathsf{N} = 4$ is chosen for easy understanding.
As stated before, the initial state of the system $x^0$ can be seeded by the decimal part $t$ of the current time.
For example, if the current time in seconds since the Epoch is 1237632934.484088,
so $t = 484088$, then $x^0 = t \text{ ($mod$ 16)}$ in binary digits, \emph{i.e.}, $x^0 = ( 0, 1, 0, 0)$.

To compute $m$ sequence, Equation~\ref{Formula} can be adapted to this example as follows:
\begin{equation}
\label{m1 fuction}
m^n=g_1(y^n)=
\left\{
\begin{array}{llccccc}
0 & \text{ if }&0 &\leqslant&\frac{y^n}{2^{32}}&<&\frac{1}{16},\\
1 & \text{ if }&\frac{1}{16} &\leqslant&\frac{y^n}{2^{32}}&<&\frac{5}{16} ,\\
2 & \text{ if }&\frac{5}{16} &\leqslant&\frac{y^n}{2^{32}}&<&\frac{11}{16},\\
3 & \text{ if }&\frac{11}{16} &\leqslant&\frac{y^n}{2^{32}}&<&\frac{15}{16},\\
4 & \text{ if }&\frac{15}{16} &\leqslant&\frac{y^n}{2^{32}}&<&1,\\
\end{array}
\right.
\end{equation}

\noindent where $y$ is generated by XORshift seeded with the current time. We can see that the probabilities of occurrences of $m=0$, $m=1$, $m=2$, $m=3$, $m=4$, are $\frac{1}{16}$, $\frac{4}{16}$, $\frac{6}{16}$, $\frac{4}{16}$, $\frac{1}{16}$, respectively. This $m$ determines what will be the next output $x$. For instance,
\begin{itemize}
\item If $m=0$, the following $x$ will be $( 0, 1, 0, 0)$.
\item If $m=1$, the following $x$ can be $( 1, 1, 0, 0)$, $( 0, 0, 0, 0)$, $( 0, 1, 1, 0)$, or $( 0, 1, 0, 1)$.
\item If $m=2$, the following $x$ can be $( 1, 0, 0, 0)$, $( 1, 1, 1, 0)$, $( 1, 1, 0, 1)$, $( 0, 0, 1, 0)$, $( 0, 0, 0, 1)$, or $( 0, 1, 1, 1)$.
\item If $m=3$, the following $x$ can be $( 0, 0, 1, 1)$, $( 1, 1, 1, 1)$, $( 1, 0, 0, 1)$, or $( 1, 0, 1, 0)$.
\item If $m=4$, the following $x$ will be $( 1, 0, 1, 1)$.
\end{itemize}

In this simulation, $m = 0, 4, 2, 2, 3, 4, 1, 1, 2, 3, 0, 1, 4,...$ Additionally, $b$ is computed with a XORshift generator too, but with another seed. We have found $b = 1, 4, 2, 2, 3, 3, 4, 1, 1, 4, 3, 2, 1,...$

Chaotic iterations are made with initial state $x^0$, vectorial logical negation $f_0$, and
strategy $S$. The result is presented in Table~\ref{table application example}. Let us recall that sequence $m$ gives the states $x^n$ to return, which are here $x^0, x^{0+4}, x^{0+4+2}, \hdots$ So, in this example, the output of the generator is: 10100111101111110011... or 4,4,11,8,1...

\begin{table*}[!t]
\centering
\begin{tabular}{|c|cc|cccccc|ccc|cccc|}
\hline
$m$ &0 & &4 & & & & & &2& &&2&&  &  \\ \hline
$k$ &0 & &4 & & &$+1$ & & &2& &&2&$+1$&  &  \\ \hline
$b$  &  & &1 &4&2&\underline{2}       &3& &3&4&&1&\underline{1}      &4&\\ \hline
$d$  &r  & &r~$\left(\begin{array}{c}1\\0\\0\\0\end{array}\right)$ & $\left(\begin{array}{c}1\\0\\0\\1\end{array}\right)$ & $\left(\begin{array}{c}1\\1\\0\\1\end{array}\right)$ & & $\left(\begin{array}{c}1\\1\\1\\1\end{array}\right)$ && r~$\left(\begin{array}{c}0\\0\\1\\0\end{array}\right)$ &$\left(\begin{array}{c}0\\0\\1\\1\end{array}\right)$ &&r~$\left(\begin{array}{c}1\\0\\0\\0\end{array}\right)$ & &$\left(\begin{array}{c}1\\0\\0\\1\end{array}\right)$  &  \\ \hline
$S$  &  & &1 &4&2&        &3& &3&4&&1& &4 &  \\ \hline
$x^{0}$ &  &$x^{0}$ & & &  
&  & &$x^{4}$ & & &   
$x^{6}$& & &&$x^{8}$  \\
0 & &0 &$\xrightarrow{1} 1$ & &
 & &   &1   & & &
1 &$\xrightarrow{1} 0$ & & & 0\\
1 &  &1 &   &   &
$\xrightarrow{2} 0$ & & &0 & & &
0 & &  &&0\\
0 & &0 & & &
 & &$\xrightarrow{3} 1$ &1 &$\xrightarrow{3} 0$ & &
0 &   & & &0  \\
0 & &0  & &$\xrightarrow{4} 1$ &
 & & &1 & &$\xrightarrow{4} 0$ &
0 & & &$\xrightarrow{4} 1$&1 \\
\hline
\end{tabular}\\
\vspace{0.5cm}
Binary Output: $x_1^{0}x_2^{0}x_3^{0}x_4^{0}x_1^{4}x_2^{4}x_3^{4}x_4^{4}x_1^{6}x_2^{6}... = 0100101110000001...$\\
Integer Output:
$x^{0},x^{4},x^{6},x^{8}... = 4,11,8,1...$
\caption{Example of New CI(XORshift,XORshift) generation}
\label{table application example}
\end{table*}

\section{Security Analysis}
\label{Security analysis}
PRNG should be sensitive with respect to the secret key and its size. Here, chaotic properties are also in close relation with the security.

\subsection{Key Space}

The PRNG proposed in this paper is based on discrete chaotic iterations. It has an initial
value $x^0\in \mathds{B}^{\mathsf{N}}$. Considering this set of initial values alone, the key space size
is equal to $2^\mathsf{N}$. In addition, this new generator combines digits of two other PRNGs. We used two different XORshifts here. Let $k$ be the key space of XORshift, so the total key space
size is close to $2^\mathsf{N}\cdot k^2$. Lastly, the impact of Equation~\ref{Formula}, in which is defined the $(m^n)$ sequence with a selector function $g_1$, must be
taken into account. This leads to conclude that the key space size is large enough to withstand
attacks.

Let us notice, to conclude this subsection, that our PRNG can use any reasonable function as selector. In this paper, $g_1()$ and $g_2()$ are adopted for demonstration purposes, where:
\begin{equation}
m^n = g_2(y^n)=
\left\{
\begin{array}{l}
\mathsf{N} \text{ if }0 \leqslant\frac{y^n}{2^{32}}<\frac{C^0_\mathsf{N}}{2^\mathsf{N}},\\
\mathsf{N}-1 \text{ if }\frac{C^0_N}{2^\mathsf{N}} \leqslant\frac{y^n}{2^{32}}<\sum_{i=0}^1\frac{C^i_\mathsf{N}}{2^\mathsf{N}},\\
\mathsf{N}-2 \text{ if }\sum_{i=0}^1\frac{C^i_\mathsf{N}}{2^\mathsf{N}} \leqslant\frac{y^n}{2^{32}}<\sum_{i=0}^2\frac{C^i_\mathsf{N}}{2^\mathsf{N}},\\
\vdots~~~~~ ~~\vdots~~~ ~~~~\\
0 \text{ if }\sum_{i=0}^{\mathsf{N}-1}\frac{C^i_\mathsf{N}}{2^\mathsf{N}} \leqslant\frac{y^n}{2^{32}}<1.\\
\end{array}
\right.
\end{equation}

We will show later that both of them can pass all of the performed tests.

\subsection{Key Sensitivity}
\begin{figure}
\centering
\includegraphics[width=3.5in]{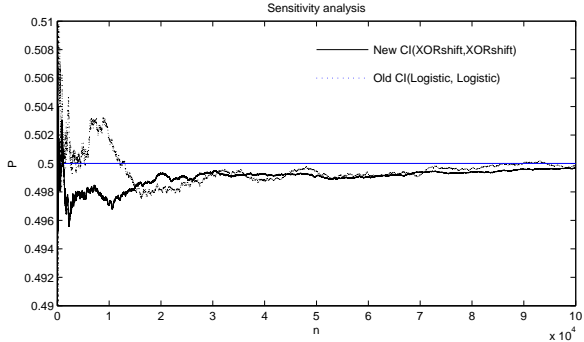}
\DeclareGraphicsExtensions.
\caption{Sensitivity analysis}
\label{Sensitivity analysis}
\end{figure}

As a consequence of its chaotic property, this PRNG is highly sensitive to the initial conditions. To illustrate this fact, several initial values are put into the chaotic system. Let $H$ be the number
of differences between the sequences obtained in this way. Suppose $n$ is the length of these
sequences. Then the variance ratio $P$, defined by $P = H / n$, is computed. The results are
shown in Figure~\ref{Sensitivity analysis} ($x$ axis is sequence lengths, $y$ axis is variance ratio $P$). For the two PRNGs, variance
ratios approach $0.50$, which indicates that the system is extremely sensitive to the initial
conditions.

\subsection{Linear Complexity}

\begin{figure}
\centering
\includegraphics[width=3.5in]{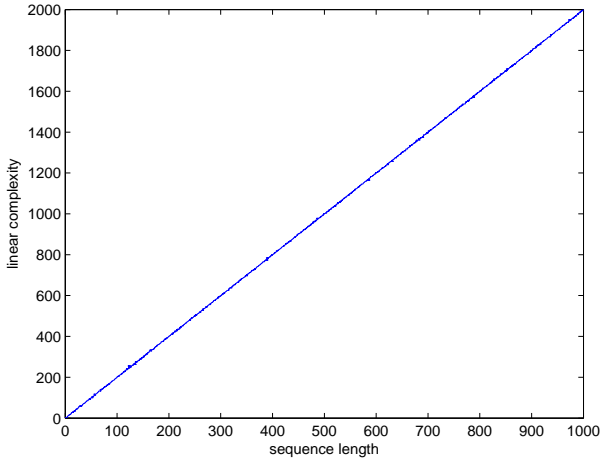}
\DeclareGraphicsExtensions.
\caption{Linear complexity}
\label{Linear complexity}
\end{figure}

The linear complexity (LC) of a sequence is the size in bits of the shortest linear feedback shift register (LFSR) which can produce this sequence. This value measures the difficulty of generating -- and perhaps analyzing -- a particular sequence.
Indeed, the randomness of a given sequence can be linked to the size of the smallest program that can produce it. LC is the size required by a LFSR to be able to produce the given sequence. The Berlekamp-Massey algorithm can measure this LC, which can be used to evaluate the security of a pseudo-random sequence.
It can be seen in Figure~\ref{Linear complexity} that the LC curve of a sample sequence of 2000 bits is close to the ideal line $C_i=i/2$, which implies that the generator has high linear complexity.

\subsection{Devaney's Chaos Property}

Generally speaking, the quality of a PRNG depends, to a large extent, on the following criteria: randomness, uniformity, independence, storage efficiency, and reproducibility. A chaotic sequence may satisfy these requirements and also other chaotic properties, as ergodicity, entropy, and expansivity. A chaotic sequence is extremely sensitive to the initial conditions. That is, even a minute difference in the initial state of the system can lead to enormous differences in the final state, even over fairly small timescales. Therefore, chaotic sequence fits the requirements of pseudo-random sequence well. Contrary to XORshift, our generator possesses these chaotic properties~\cite{guyeux09},\cite{wang2009}.
However, despite a large number of papers published in the field of chaos-based pseudo-random generators, the impact of this research is rather marginal. This is due to the following reasons: almost all PRNG algorithms using chaos are based on dynamical systems defined on continuous sets (\emph{e.g.}, the set of real numbers). So these generators are usually slow, requiring considerably more storage space, and lose their chaotic properties during computations as mentioned earlier in this paper. These major problems restrict their use as generators~\cite{Kocarev2001}.

In this paper, we do not simply integrate chaotic maps hoping that the implemented algorithm remains chaotic. Indeed, the PRNG we conceive is just discrete chaotic iterations and we have proven in \cite{guyeux09} that these iterations produce a topological chaos as defined by Devaney: they are regular, transitive, and sensitive to initial conditions. This famous definition of a chaotic behavior for a dynamical system implies unpredictability, mixture, sensitivity, and uniform repartition. Moreover, as only integers are manipulated in discrete chaotic iterations, the chaotic behavior of the system is preserved during computations, and these computations are fast.

Let us now explore the topological properties of our generator and their consequences concerning the quality of the generated pseudo-random sequences.

\subsection{Topological Consequences}

We have proven in \cite{gfb10:ip} that chaotic iterations are expansive and topologically mixing. These topological properties are inherited by the generators we presented here. In particular, any error on the seed are magnified until being equal to the constant of expansivity.
We will now investigate the consequences of being chaotic, as defined by Devaney. 

First of all, the transitivity property implies the indecomposability of the system:

\begin{definition}
A dynamical system $\left( \mathcal{X}, f\right)$ is indecomposable if it is not the union of two closed sets $A, B \subset \mathcal{X}$ such that $f(A) \subset A, f(B) \subset B$.
\end{definition}

Thus it is impossible to reduce the set of the outputs generated by our PRNG, in order to reduce its complexity. Moreover, it is possible to show that Old and New CI generators are strongly transitive:

\begin{definition}
A dynamical system $\left( \mathcal{X}, f\right)$ is strongly transitive if $\forall x,y \in \mathcal{X},$ $\forall r > 0,$ $\exists z \in \mathcal{X},$ $d(z,x) \leqslant r \Rightarrow$ $\exists n \in \mathds{N}^*,$ $f^n(z)=y$.
\end{definition}

In other words, for all $x,y \in \mathcal{X}$, it is possible to find a point $z$ in the neighborhood of $x$ such that an iterate $f^n(z)$ is $y$. Indeed, this result has been established during the proof of the transitivity presented in~\cite{guyeux09}. Among other things, the strong transitivity property leads to the fact that without the knowledge of the seed, all of the outputs are possible. Additionally, no point of the output space can be discarded when studying our PRNG: it is intrinsically complicated and it cannot be simplified.

Finally, these generators possess the instability property:

\begin{definition}
A dynamical system $\left( \mathcal{X}, f\right)$ is unstable if for all $x \in \mathcal{X}$, the orbit $\gamma_x:n \in \mathds{N} \longmapsto f^n(x)$ is unstable, that is: $\exists \varepsilon > 0,$ $\forall \delta > 0,$ $\exists y \in \mathcal{X},$ $\exists n \in \mathds{N},$ $d(x,y) < \delta$ and $d\left(\gamma_x(n), \gamma_y(n)\right) \geqslant \varepsilon.$
\end{definition}

This property, which is implied by the sensitive dependence to the initial condition, leads to the fact that in all of the neighborhoods of any $x$, there are points that are separate from $x$ under iterations of $f$. We thus can claim that the behavior of our generators is unstable.

\section{Statistical Analysis}
\label{Comparative analysis}
\subsection{Basic Common Tests}

\subsubsection{Comparative test parameters}
In this section, five well-known statistical tests~\cite{Menezes1997} are used as comparison tools. They encompass frequency and autocorrelation tests. In what follows, $s = s^0,s^1,s^2,\dots , s^{n-1}$ denotes a binary sequence of length $n$. The question is to determine whether this sequence possesses some specific characteristics that a truly random sequence would be likely to exhibit. The tests are introduced in this subsection and results are given in the next one.

\paragraph{Frequency test (monobit test)}
The purpose of this test is to check if the numbers of 0's and 1's are approximately equal in $s$, as it would be expected for a random sequence. Let $n_0, n_1$ denote these numbers. The statistic used here is:
\begin{equation*}
 X_1=\frac{(n_0-n_1)^2}{n}, 
\end{equation*}
\noindent which approximately follows a $\chi^2$ distribution with one degree of freedom when $n\geqslant 10^7$.

\paragraph{Serial test (2-bit test)}
The purpose of this test is to determine if the number of occurrences of 00, 01, 10, and 11 as subsequences of $s$ are approximately the same. Let $n_{00} , n_{01} ,n_{10}$, and $n_{11}$ denote the number of occurrences of $00, 01, 10$, and $11$ respectively. Note that $n_{00} + n_{01} + n_{10} + n_{11} = n-1$ since the subsequences are allowed to overlap. The
statistic used here is:
\begin{equation*}
X_2=\frac{4}{n-1}(n_{00}^2+n_{01}^2+n_{10}^2+n_{11}^2)-\frac{2}{n}(n_0^2+n_1^2)+1,
\end{equation*}
\noindent which approximately follows a $\chi^2$ distribution with 2 degrees of freedom if $n\geqslant 21$.

\paragraph{Poker test}
The poker test studies if each pattern of length $m$ (without overlapping) appears the same number of times in $s$. Let $\lfloor \frac{n}{m} \rfloor\geqslant 5 \times 2^m$ and $k= \lfloor \frac{n}{m} \rfloor $. Divide the sequence $s$ into $k$ non-overlapping parts, each of length $m$. Let $n_i$ be the number of occurrences of the $i^{th}$ type of sequence of length $m$, where $1 \leqslant i \leqslant 2^m$. The statistic used is
\begin{equation*}
X_3=\dfrac{2^m}{k}\left(\displaystyle{\sum^{2^m}_{i=1}n^2_i}\right)-k,
\end{equation*}
which approximately follows a $\chi^2$ distribution with $2^m-1$ degrees of freedom. Note that the poker test is a generalization of the frequency test: setting $m = 1$ in the poker test yields the frequency test.

\paragraph{Runs test}
The purpose of the runs test is to figure out whether the number of runs of various lengths in the sequence $s$ is as expected for a random sequence. A run is defined as a pattern of all zeros or all ones, a block is a run of ones, and a gap is a run of zeros. The expected number of gaps (or blocks) of length $i$ in a random sequence of length $n$ is $e_i = \frac{n-i+3}{2^{i+2}}$. Let $k$ be equal to the largest integer $i$ such that $e_i \geqslant 5$. Let
$B_i , G_i$ be the number of blocks and gaps of length $i$ in $s$, for each $i \in \llbracket 1, k\rrbracket$. The statistic used here will then be:
\begin{equation*}
\displaystyle{X_4=\sum^k_{i=1}\frac{(B_i-e_i)^2}{e_i}+\sum^k_{i=1}\frac{(G_i-e_i)^2}{e_i}},
\end{equation*}
\noindent which approximately follows a $\chi^2$ distribution with $2k - 2$ degrees of freedom.

\paragraph{Autocorrelation test}

The purpose of this test is to check for coincidences between the sequence $s$ and (non-cyclic) shifted versions of it. Let $d$ be a fixed integer, $ 1 \leqslant d \leqslant \lfloor n/2 \rfloor$. The value $A(d) = \sum_{i=0}^{n-d-1} s_i\oplus s_{i+d}$ is the amount of bits not equal between the sequence and itself displaced by $d$ bits. The statistic used here is:\newline
\begin{equation*}
X_5=|2(A(d)-\frac{n-d}{2})/\sqrt{n-d}|,
\end{equation*}
\noindent which approximately follows a normal distribution $\mathcal{N}(0, 1)$ if $n-d \geqslant 10$. Since small values of $A(d)$ are as unexpected as large values, a two-sided test should be used.

\subsubsection{Comparison}
\begin{table*}[!t]
\renewcommand{\arraystretch}{1.3}
\caption{Comparison with Old CI(Logistic, Logistic) for a $2 \times 10^5$ bits sequence}
\label{Comparison2}
\centering
\begin{tabular}{ccccccc}
\hline
Method & Monobit ($X_1$)& Serial ($X_2$)& Poker ($X_3$)& Runs ($X_4$)& Autocorrelation ($X_5$)& Time \\ \hline
Logistic map &0.1280&0.1302&240.2893&26.5667&0.0373&0.965s \\
XORshift &1.7053&2.1466&248.9318&18.0087&0.5009&0.096s \\
Old CI(Logistic, Logistic) &1.0765&1.0796&258.1069&20.9272&1.6994&0.389s \\
New CI(XORshift,XORshift) &0.3328&0.7441&262.8173&16.7877&0.0805&0.197s\\
\hline
\end{tabular}
\end{table*}
We show in Table~\ref{Comparison2} a comparison among our new generator CI(XORshift, XORshift), its old version denoted Old CI(Logistic, Logistic), a basic PRNG based on logistic map, and a simple XORshift. In this table, time (in seconds) is related to the duration needed by each algorithm to generate a $2 \times 10^5$ bits long sequence. The test has been conducted using the same computer and compiler with the same optimization settings for both algorithms, in order to make the test as fair as possible. The results confirm that the proposed generator is a lot faster than the old one, while the statistical results are better for most of the parameters, leading to the conclusion that the new PRNG is more secure than the old one. Although the logistic map also has good results, it is too slow to be implemented in Internet applications, and this map is known to present various bias leading to severe security issues.

\begin{figure}
\centering
\includegraphics[width=3.5in,height=2in]{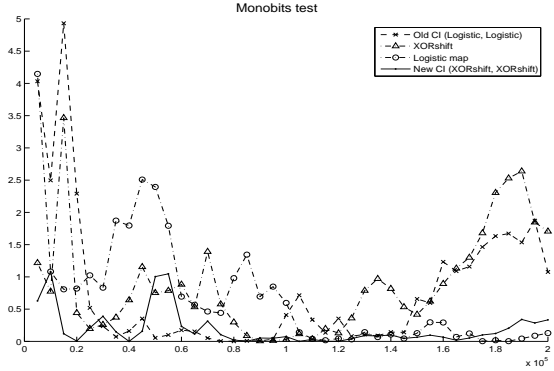}
\caption{Comparison of monobits tests}
\label{monobits}
\end{figure}

\begin{figure}
\centering
\includegraphics[width=3.5in,height=2in]{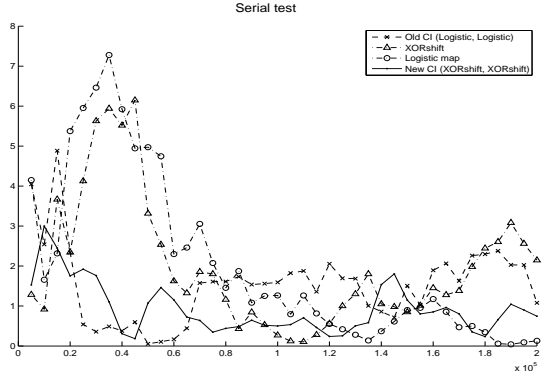}
\caption{Comparison of serial tests}
\label{serial}
\end{figure}

\begin{figure}
\centering
\includegraphics[width=3.5in,height=2in]{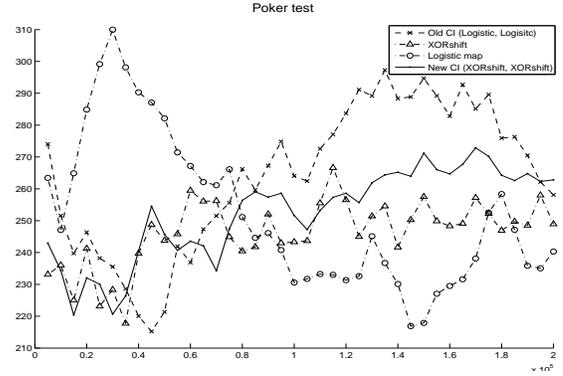}
\caption{Comparison of poker tests}
\label{poker}
\end{figure}

\begin{figure}
\centering
\includegraphics[width=3.5in,height=2in]{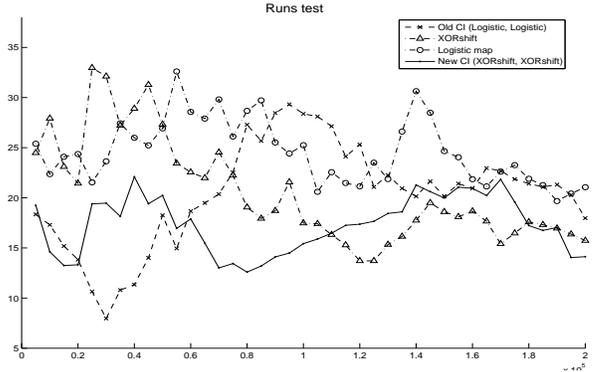}
\caption{Comparison of runs tests}
\label{runs}
\end{figure}

\begin{figure}
\centering
\includegraphics[width=3.5in,height=2in]{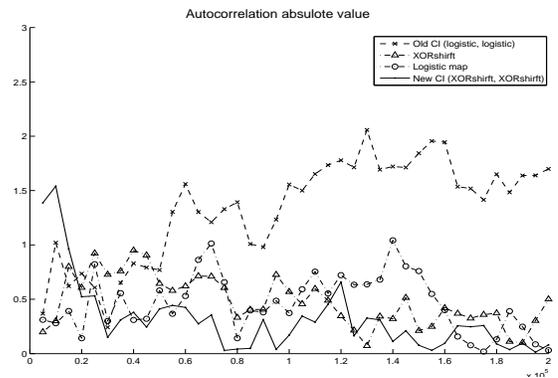}
\caption{Comparison of autocorrelation tests}
\label{autocorrelation}
\end{figure}

As a comparison of the overall stability of these PRNGs, similar tests have been computed for different sequence lengths (see Figures \ref{monobits} - \ref{autocorrelation}).
For the monobit test comparison (Figure \ref{monobits}), almost all of the PRNGs present the same issue: the beginning values are a little high. However, for our new generator, the values are stable in a low level which never exceeds 1.2. Indeed, the new generator distributes very randomly the zeros and ones, whatever the length of the desired sequence. 
It can also be remarked that the old generator presents the second best performance, due to its use of chaotic iterations.

Figure \ref{serial} shows the serial test comparison. The new generator outperforms this test, but the score of the old generator is not bad either: their occurrences of 00, 01, 10, and 11 are very close to each other.

The poker test comparison with $m=8$ is shown in Figure \ref{poker}. XORshift is the most stable generator in all of these tests, and the logistic map also becomes good when producing sequences of length greater than $1 \times 10^5$. 
Our old and new generators present a similar trend, with a maximum in the neighborhood of $1.7 \times 10^5$. These scores are not so good, even though the new generator has a better behavior than the old one. 
Indeed, the value of $m$ and the length of the sequences should be enlarged to be certain that the chaotic iterations express totally their complex behavior. In that situation, the performances of our generators in the poker test can be improved.

The graph of the new generator is the most stable one during the runs test comparison (Figure \ref{runs}). Moreover, this trend is reinforced when the lengths of the tested sequences are increased.

The comparison of autocorrelation tests is presented in Figure \ref{autocorrelation}. The new generator clearly dominates these tests, whereas the score of the old generator is surprisingly bad. This difference between two generators based on chaotic iterations can be explained by the fact that the improvements realized to define the new generator lead to a more randomly output.

To sum up we can claim that the new generator, which is faster than its former version, outperforms all of the other generators in these statistical tests, especially when producing long output sequences.

\subsection{NIST Statistical Test Suite}

\subsubsection{Presentation}

Among the numerous standard tests for pseudo-randomness, a convincing way to prove the quality of the produced sequences is to confront them with the NIST (National Institute of Standards and Technology) Statistical Test Suite SP 800-22, 
released by the Information Technology Laboratory in August 25, 2008.

The NIST test suite, SP 800-22, is a statistical package consisting of 15 tests. They were developed to measure the randomness of (arbitrarily long) binary sequences produced by either hardware or software based cryptographic pseudorandom number generators. These tests focus on a variety of different types of non-randomness that could occur in such sequences. These 15 tests include in the NIST test suite are described in the Appendix.

\subsubsection{Interpretation of empirical results}

$\mathbb{P}$ is the ``tail probability'' that the chosen test statistic will assume values that are equal to or worse than the observed test statistic value when considering the null hypothesis.
For each statistical test, a set of $\mathbb{P}$s is produced from a set of sequences obtained by our generator (i.e., 100 sequences are generated and tested, hence 100 $\mathbb{P}$s are produced). 

Empirical results can be interpreted in various ways. In this paper, 
we check whether the $\mathbb{P}$s are uniformly distributed, via an application of
a $\chi^2$ distribution and the determination of a $\mathbb{P}_T$
corresponding to the Goodness-of-Fit distributional test
on the $\mathbb{P}$s obtained for an arbitrary statistical test. 

%
%
%
%
%
If $\mathbb{P}_T \geq 0.0001$, then the sequences can be considered to be uniformly distributed.
In our experiments, 100 sequences (s = 100) of 1,000,000 bits are generated and tested. If the value $\mathbb{P}_T$ of a least one test is smaller than 0.0001, the sequences are considered to be not good enough and the generator is unsuitable.

Table~\ref{The passing rate} shows $\mathbb{P}_T$ for the sequences based on discrete chaotic iterations using different schemes. If there are at least two statistical values in a test, this test is marked with an asterisk and the average is computed to characterize the statistical values.

We can conclude from Table \ref{The passing rate} that the worst situations are obtained with the New CI ($m^n=y^n~mod~N$) and New CI (no mark) generators. 
Old CI, New CI ($m^n=g_1(y^n)$), and New CI ($m^n=g_2(y^n)$) have successfully passed the NIST statistical test suite. These results and the conclusion obtained from the aforementioned basic tests reinforce the confidence that can be put in the good behavior of chaotic CI PRNGs, thus making them suitable for security applications as information hiding and digital watermarking.

\section{Application Example in Information Hiding}
\label{An application example of the proposed PRNG}

\subsection{Introduction}

Information hiding is now an integral part of Internet technologies. In the field of social search engines, for example, contents like pictures or movies are tagged with descriptive labels by contributors, and search results are determined by these descriptions. These collaborative taggings, used for example in Flickr~\cite{Frick} and Delicious~\cite{Delicious} websites, contribute to the development of a Semantic Web, in which any Web page contains machine-readable metadata that describe its content. Information hiding technologies can be used for embedding these metadata. The advantage of its use is the possibility to realize  social search without websites and databases: descriptions are directly embedded into media, whatever their formats. Robustness is required in this situation, as descriptions should resist to modifications like resizing, compression, and format conversion.

The Internet security field is also concerned by watermarking technologies. Steganography and cryptography are supposed to be used by terrorists to communicate through the Internet. Furthermore, in the areas of defense or in industrial espionage, many information leaks using steganographic techniques have been reported. Lastly, watermarking is often cited as a possible solution to digital rights managements issues, to counteract piracy of digital work in an Internet based entertainment world~\cite{Nakashima2003}.

\subsection{Definition of a Chaos-Based Information Hiding Scheme}
\label{sec:Algo}

Let us now introduce our information hiding scheme based on CI generator.

\subsubsection{Most and least significant coefficients}

Let us define the notions of most and least significant coefficients of an image.

\begin{Definition}
\label{definitionMSC}
For a given image, most significant coefficients (in short MSCs), are coefficients that allow the description of the relevant part of the image, \emph{i.e.}, its richest part (in terms of embedding information), through a sequence of bits.
\end{Definition}

\begin{figure}[htb]

\begin{minipage}[b]{1.0\linewidth}
  \centering
 \centerline{\epsfig{figure=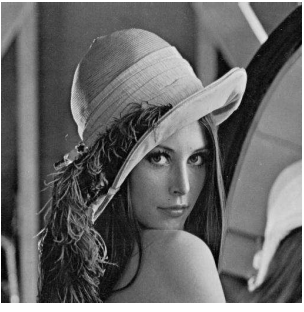,width=4cm}}
  \centerline{(a) Lena.}
\end{minipage}

\begin{minipage}[b]{.48\linewidth}
  \centering
 \centerline{\epsfig{figure=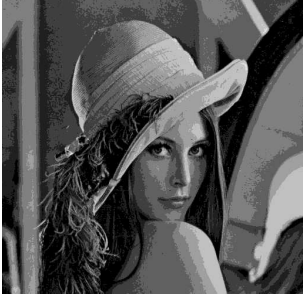,width=4cm}}
  \centerline{(b) MSCs of Lena.}
\end{minipage}
\hfill
\begin{minipage}[b]{0.48\linewidth}
  \centering
 \centerline{\epsfig{figure=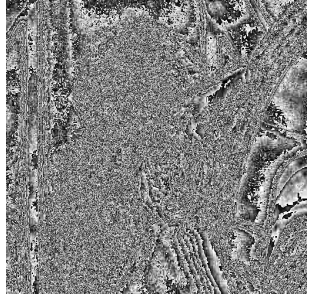,width=4cm}}
  \centerline{(c) LSCs of Lena ($\times 17$).}
\end{minipage}
\caption{Example of most and least significant coefficients of Lena.}
\label{fig:MSCLC}
\end{figure}

For example, in a spatial description of a grayscale image, a definition of MSCs can be the sequence constituted by the first four bits of each pixel (see Figure~\ref{fig:MSCLC}). In a discrete cosine frequency domain description, each $8\times 8$ block of the carrier image is mapped onto a list of 64 coefficients. The energy of the image is mostly contained in a determined part of themselves, which can constitute a possible sequence of MSCs.

\begin{Definition}
\label{definitionLSC}
By least significant coefficients (LSCs), we mean a translation of some insignificant parts of a medium in a sequence of bits (insignificant can be understand as: ``which can be altered without sensitive damages'').
\end{Definition}

These LSCs can be, for example, the last three bits of the gray level of each pixel (see Figure~\ref{fig:MSCLC}). Discrete cosine, Fourier, and wavelet transforms can be used also to generate LSCs and MSCs. Moreover, these definitions can be extended to other types of media.

LSCs are used during the embedding stage. Indeed, some of the least significant coefficients of the carrier image will be chaotically chosen by using our PRNG. These bits will be either switched or replaced by the bits of the watermark. The MSCs are only useful in case of authentication; mixture and embedding stages depend on them. Hence, a coefficient should not be defined at the same time as a MSC and a LSC: the last can be altered while the first is needed to extract the watermark.

\subsubsection{Stages of the scheme}

Our CI generator-based information hiding scheme consists of two stages: (1) mixture of the watermark and (2) its embedding.

\paragraph{Watermark mixture}

Firstly, for security reasons, the watermark can be mixed before its embedding into the image. A first way to achieve this stage is to apply the bitwise exclusive or (XOR) between the watermark and the New CI generator. In this paper, we introduce a new mixture scheme based on chaotic iterations. Its chaotic strategy, which depends on our PRNG, will be highly sensitive to the MSCs, in the case of an authenticated watermarking.

\paragraph{Watermark embedding}

Some LSCs will be switched, or substituted by the bits of the possibly mixed watermark. To choose the sequence of LSCs to be altered, a number of integers, less than or equal to the number $\mathsf{M}$ of LSCs corresponding to a chaotic sequence $U$, is generated from the chaotic strategy used in the mixture stage. Thus, the $U^{k}$-th least significant coefficient of the carrier image is either switched, or substituted by the $k^{th}$ bit of the possibly mixed watermark. In case of authentication, such a procedure leads to a choice of the LSCs that are highly dependent on the MSCs~\cite{guyeux10}.

On the one hand, when the switch is chosen, the watermarked image is obtained from the original image whose LSBs $L = \mathds{B}^{\mathsf{M}}$ are replaced by the result of some chaotic iterations. Here, the iterate function is the vectorial Boolean negation,
\begin{equation}
f_0:(x_1,...,x_\mathsf{M}) \in \mathds{B}^\mathsf{M} \longmapsto (\overline{x_1},...,\overline{x_\mathsf{M}}) \in \mathds{B}^\mathsf{M},
\end{equation}
the initial state is $L$, and the strategy is equal to $U$. In this case, the whole embedding stage satisfies the topological chaos properties~\cite{guyeux10}, but the original medium is required to extract the watermark. On the other hand, when the selected LSCs are substituted by the watermark, its extraction can be done without the original cover (blind watermarking). In this case, the selection of LSBs still remains chaotic because of the use of the New CI generator, but the whole process does not satisfy topological chaos~\cite{guyeux10}. The use of chaotic iterations is reduced to the mixture of the watermark. See the following sections for more detail.

\paragraph{Extraction}

The chaotic strategy can be regenerated even in the case of an authenticated watermarking, because the MSCs have not changed during the embedding stage. Thus, the few altered LSCs can be found, the mixed watermark can be rebuilt, and the original watermark can be obtained. In case of a switch, the result of the previous chaotic iterations on the watermarked image should be the original cover. The probability of being watermarked decreases when the number of differences increase.

If the watermarked image is attacked, then the MSCs will change. Consequently, in case of authentication and due to the high sensitivity of our PRNG, the LSCs designed to receive the watermark will be completely different. Hence, the result of the recovery will have no similarity with the original watermark.

\begin{figure}[htb]
\centerline{\epsfig{figure=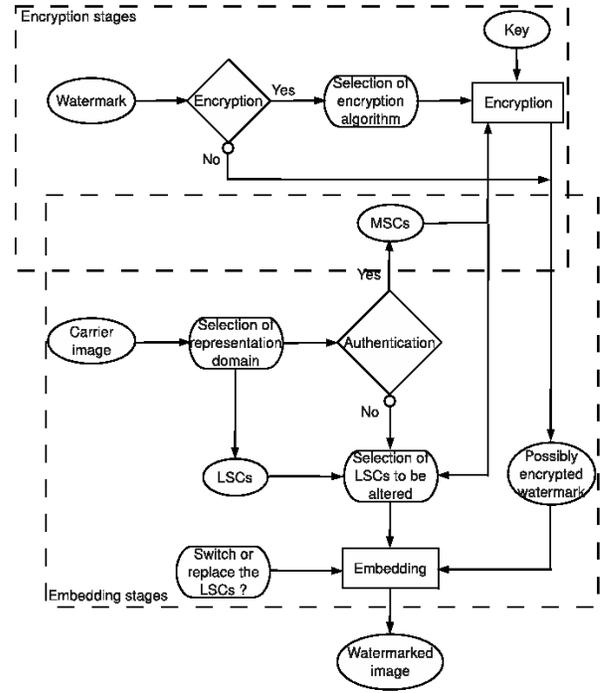,width=8.cm}}
\caption{The chaos-based data hiding decision tree.}
\label{fig:organigramme}
\end{figure}

The chaos-based data hiding scheme is summed up in Figure~\ref{fig:organigramme}.

\subsection{Application Example}

\subsubsection{Experimental protocol}

\begin{figure}[!t]
\centering
\subfloat [The original image]{\includegraphics[scale=0.23]{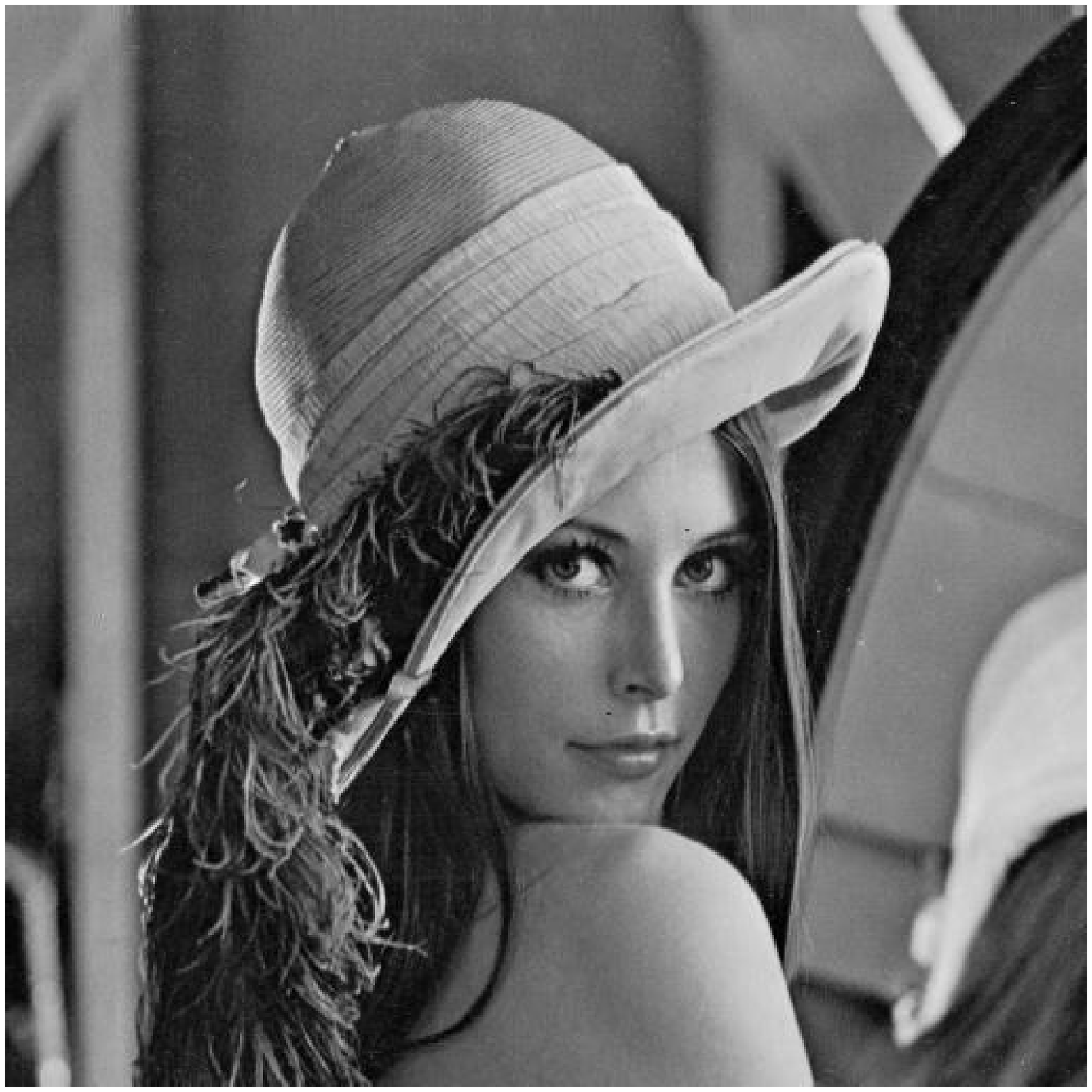}}
\hfil
\subfloat [The watermark]{\includegraphics[scale=0.4]{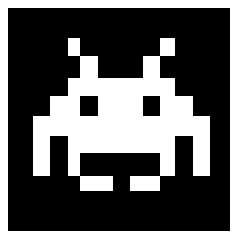}%
}
\caption{Original images}
\label{Original images}
\end{figure}

In this subsection, a concrete example is given: a watermark is encrypted and embedded into a cover image using the scheme presented in the previous section and CI(XORshift, XORshift). The carrier image is the well-known Lena, which is a 256 grayscale image, and the watermark is the $64\times 64$ pixels binary image depicted in Figure~\ref{Original images}.

\begin{figure}[!t]
\centering
\subfloat [Differences with the original]{\includegraphics[scale=0.42]{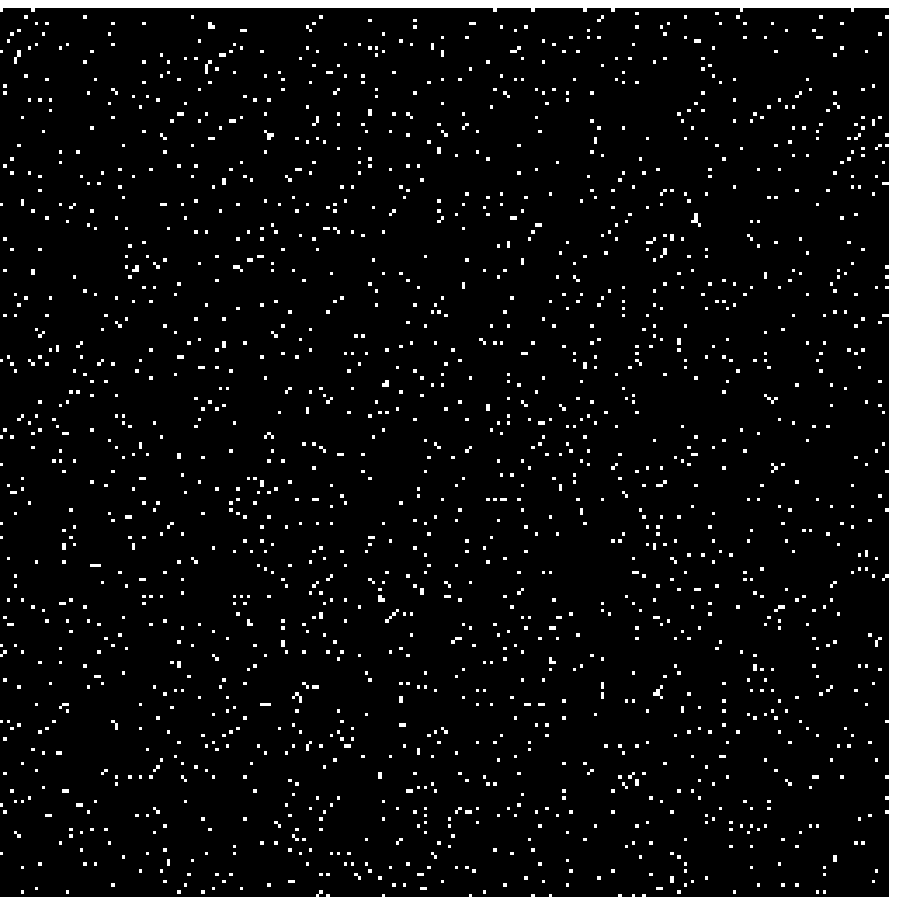}%
}
\hfil
\subfloat [Encrypted watermark]{\includegraphics[scale=0.4]{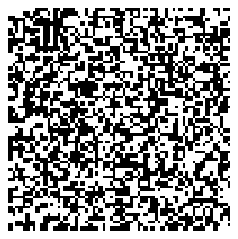}%
}
\caption{Encrypted watermark and differences}
\label{Encrypted watermark and differences}
\end{figure}

The watermark is encrypted by using chaotic iterations: the initial state $x^{0}$ is the watermark, considered as a Boolean vector, the iteration function is the vectorial logical negation, and the chaotic strategy $(S^{k})_{k\in \mathds{N}}$ is defined with CI(XORshift, XORshift), where initial parameters constitute the secret key and $N=64$. Thus, the encrypted watermark is the last Boolean vector generated by these chaotic iterations. An example of such an encryption is given in Figure~\ref{Encrypted watermark and differences}.

Let $L$ be the $256^3$ Booleans vector constituted by the three last bits of each pixel of Lena and $U^k$ defined by the sequence:
\begin{equation}
\left\{
\begin{array}{lll}
U^{0} & = & S^{0} \\
U^{n+1} & = & S^{n+1}+2\times U^{n}+n ~ [mod ~ 256^3].%
\end{array}%
\right.
\end{equation}
The watermarked Lena $I_w$ is obtained from the original Lena, whose three last bits are replaced by the result of $64^2$ chaotic iterations with initial state $L$ and strategy $U$ (see Figure~\ref{Encrypted watermark and differences}).

The extraction of the watermark can be obtained in the same way. Remark that the map $\theta \mapsto 2\theta $ of the torus, which is the famous dyadic transformation (a well-known example of topological chaos~\cite{Dev89}), has been chosen to make $(U^{k})_{k \leqslant 64^2}$ highly sensitive to the strategy. As a consequence, $(U^{k})_{k \leqslant 64^2}$ is highly sensitive to the alteration of the image: any significant modification of the watermarked image will lead to a completely different extracted watermark, thus giving a way to authenticate media through the Internet.

\begin{table*}[!t]
\renewcommand{\arraystretch}{1.3}
\caption{SP 800-22 test results ($\mathbb{P}_T$)}
\label{The passing rate}
\centering
\begin{tabular}{|l||c|c|c|c|c|}
\hline
Method & New CI ($m^n=y^n~mod~N$)& New CI (no mark)& Old CI & New CI ($g_1()$)&New CI ($g_2()$)\\ \hline\hline

Frequency (Monobit) Test 			&0.0004&0.0855	&0.595549&0.474986 &0.419\\ \hline
Frequency Test within a Block  		&0&0	&0.554420&0.897763&0.6786\\ \hline
Runs Test  					&0.2896&0.5544	&0.455937&0.816537&0.3345\\ \hline
Longest Run of Ones in a Block Test 		&0.0109&0.4372	&0.016717&0.798139&0.8831 \\ \hline
Binary Matrix Rank Test  			&0&0.6579	&0.616305&0.262249&0.7597\\ \hline
Discrete Fourier Transform (Spectral) Test 	&0&0	&0.000190&0.007160&0.0008 \\ \hline
Non-overlapping Template Matching Test* 	&0.020071&0.37333&0.532252&0.449916& 0.51879\\ \hline
Overlapping Template Matching Test	&0&0	 &0.334538&0.514124 	&0.2492\\ \hline
Maurer's ``Universal Statistical'' Test  	&0.6993&0.9642	&0.032923&0.678686&0.1296\\ \hline
Linear Complexity Test 			&0.3669&0.924	&0.401199&0.657933 &0.3504\\ \hline
Serial Test* (m=10) 			&0&0.28185&0.013396&0.425346 	&0.2549\\ \hline
Approximate Entropy Test (m=10) 	&0&0.3838	&0.137282&0.637119 	&0.7597\\ \hline
Cumulative Sums (Cusum) Test* 		&0&0	&0.046464&0.279680&0.34245\\ \hline
Random Excursions Test* 			&0.46769&0.34788&0.503622&0.287409 &0.18977\\ \hline
Random Excursions Variant Test* 		&0.28779&0.46505&0.347772&0.486686& 0.26563\\ \hline
Success 				&8/15&11/15	& 15/15 & 15/15 	&15/15\\ \hline
\hline
\end{tabular}
\end{table*}

Let us now evaluate the robustness of the proposed method.

\subsubsection{Robustness evaluation}

In what follows, the embedding domain is the spatial domain, CI(XORshift,XORshift) has been used to encrypt the watermark, MSCs are the four first bits of each pixel (useful only in case of authentication), and LSCs are the three next bits.

To prove the efficiency and the robustness of the proposed algorithm, some
attacks are applied to our chaotic watermarked image. For each attack, a
similarity percentage with the watermark is computed, this percentage is the
number of equal bits between the original and the extracted watermark, shown
as a percentage. Let us notice that a result less than or equal to $50\%$
implies that the image has probably not been watermarked.

\paragraph{Zeroing attack}

\begin{figure}[htb]
\begin{minipage}[b]{.48\linewidth}
  \centering
 \centerline{\epsfig{figure=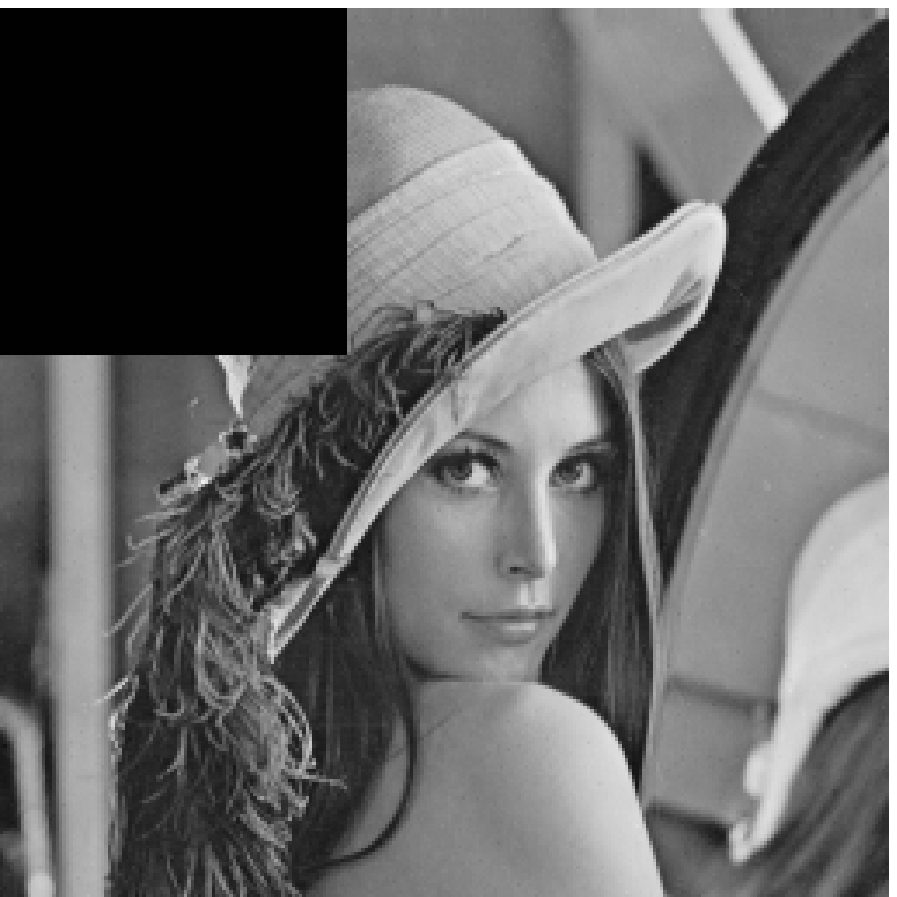,width=3.3cm}}
  \centerline{(a) Cropping attack}
\end{minipage}
\hfill
\begin{minipage}[b]{0.48\linewidth}
  \centering
 \centerline{\epsfig{figure=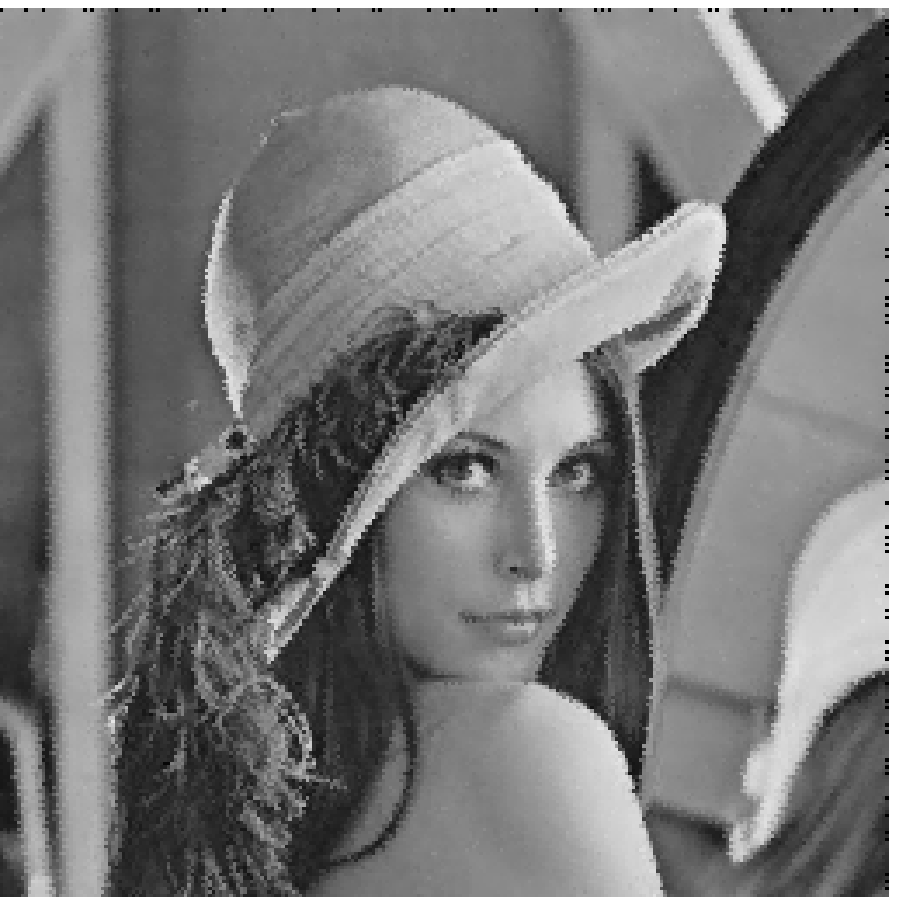,width=3.3cm}}
  \centerline{(b) Rotation attack}
\end{minipage}
\caption{Watermarked Lena after attacks.}
\label{fig:LenaAttack}
\end{figure}

In this kind of attack, a watermarked image is zeroed, such as in Figure \ref{fig:LenaAttack}(a). In this case, the results in Table 1 have been obtained.

\begin{center}
\begin{footnotesize}
\begin{tabular}{|c|c||c|c|}
\hline
\multicolumn{2}{|c||}{UNAUTHENTICATION}  & \multicolumn{2}{c|}{AUTHENTICATION}\\ 
\hline
Size (pixels) & Similarity & Size (pixels) & Similarity \\
 \hline
10 & 99.08\% & 10 & 91.77\% \\
50 & 97.31\% & 50 & 55.43\% \\
100 & 92.43\% & 100 & 51.52\% \\
200 & 70.75\% & 200 & 50.60\% \\
\hline
\end{tabular}
\end{footnotesize}\\
\vspace{0.5cm}
\textbf{Table. 1}. ~Cropping attacks
\end{center}

\begin{figure}[htb]
\begin{minipage}[b]{1.0\linewidth}
  \centering
 \centerline{\epsfig{figure=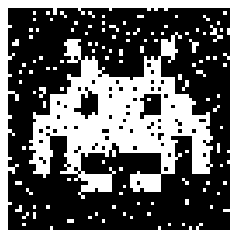,width=2cm}}
  \centerline{(a) Unauthentication ($50\times 50$).}
\end{minipage}
\begin{minipage}[b]{.48\linewidth}
  \centering
 \centerline{\epsfig{figure=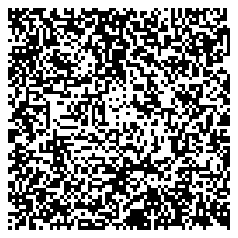,width=2cm}}
  \centerline{(b) Authentication  ($50\times 50$).}
\end{minipage}
\hfill
\begin{minipage}[b]{0.48\linewidth}
  \centering
 \centerline{\epsfig{figure=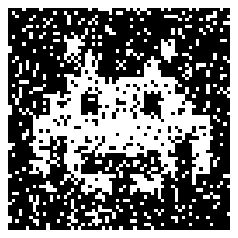,width=2cm}}
  \centerline{(c) Authentication  ($10\times 10$).}
\end{minipage}
\caption{Extracted watermark after a cropping attack.}
\label{fig:Dechiffrement_invader}
\end{figure}

In Figure \ref{fig:Dechiffrement_invader}, the decrypted watermarks are shown after a crop of 50 pixels and after a crop of 10 pixels, in the authentication case.

By analyzing the similarity percentage between the original and the
extracted watermark, we can conclude that in case of unauthentication, the
watermark still remains after a zeroing attack: the desired robustness is
reached. It can be noticed that zeroing sizes and percentages are rather
proportional.

In case of authentication, even a small change of the carrier image (a crop
by $10\times 10$ pixels) leads to a really different extracted watermark.
In this case, any attempt to alter the carrier image will be signaled, the
image is well authenticated.

\paragraph{Rotation attack}

Let $r_{\theta }$ be the rotation of angle $\theta $ around the center $%
(128, 128)$ of the carrier image. So, the transformation $r_{-\theta }\circ
r_{\theta }$ is applied to the watermarked image, which is altered as in Figure \ref{fig:LenaAttack}. The results in Table 2 have been obtained.

\begin{center}
\begin{footnotesize}
\begin{tabular}{|c|c||c|c|}
\hline
\multicolumn{2}{|c||}{UNAUTHENTICATION}  & \multicolumn{2}{c|}{AUTHENTICATION}\\ 
\hline
Angle (degree) & Similarity & Angle (degree) & Similarity \\
 \hline
2 & 96.44\% & 2 & 73.40\% \\
5 & 93.32\% & 5 & 60.56\% \\
10 & 90.68\% & 10 & 52.11\% \\
25 & 78.13\% & 25 & 51.97\% \\
\hline
\end{tabular}
\end{footnotesize}\\
\vspace{0.5cm}
\textbf{Table. 2}. ~Rotation attacks

\end{center}

The same conclusion as above can be declaimed: this watermarking method
satisfies the desired properties.

\paragraph{JPEG compression}

A JPEG compression is applied to the watermarked image, depending on a
compression level. Let us notice that this attack leads to a change of
the representation domain (from spatial to DCT domain). In this case, the results in Table 3 have been obtained.

\begin{center}
\begin{footnotesize}
\begin{tabular}{|c|c||c|c|}
\hline
\multicolumn{2}{|c||}{UNAUTHENTICATION}  & \multicolumn{2}{c|}{AUTHENTICATION}\\ 
\hline
Compression & Similarity & Compression & Similarity \\
 \hline
2 & 85.76\% & 2 & 56.42\% \\
5 & 67.62\% & 5 & 52.12\% \\
10 & 62.43\% & 10 & 48.22\% \\
20 & 54.74\% & 20 & 49.07\% \\
\hline
\end{tabular}
\end{footnotesize}\\
\vspace{0.5cm}
\textbf{Table. 3}. ~JPEG compression attacks
\end{center}

A very good authentication through JPEG attack is obtained. As for the
unauthentication case, the watermark still remains after a compression level
equal to 10. This is a good result if we take into account the fact that we
use spatial embedding.

\paragraph{Gaussian noise}

Watermarked image can be also attacked by the addition of a Gaussian noise, depending on a standard deviation. In this case, the results in Table 4 have been obtained.

\begin{center}
\begin{footnotesize}
\begin{tabular}{|c|c||c|c|}
\hline
\multicolumn{2}{|c||}{UNAUTHENTICATION}  & \multicolumn{2}{c|}{AUTHENTICATION}\\ 
\hline
Standard dev. & Similarity & Standard dev. & Similarity \\
 \hline
1 & 81.14\% & 1 & 55.57\% \\
2 & 75.01\% & 2 & 52.63\% \\
3 & 67.64\% & 3 & 52.68\% \\
5 & 57.48\% & 5 & 51.34\% \\
\hline
\end{tabular}
\end{footnotesize}\\
\vspace{0.5cm}
\textbf{Table. 4}. ~Gaussian noise attacks
\end{center}

Once again we remark that good results are obtained, especially if we keep in
mind that a spatial representation domain has been chosen.

\section{Conclusion and Future Work}
\label{Conclusions and Future Work}

In this paper, the pseudo-random generator proposed in \cite{wang2009} has been improved. By using XORshift instead of logistic map and due to a rewrite of the way to generate strategies, the generator based on chaotic iterations works faster and is more secure. The speed and randomness of this new PRNG has been compared to its former version, to XORshift, and to a generator based on logistic map. This comparison shows that CI(XORshift, XORshift) offers a sufficient speed and level of security for a wide range of Internet usages as cryptography and information hiding.

In future work, we will continue to try to improve the speed and security of this PRNG, by exploring new strategies and iteration functions. Its chaotic behavior will be deepened by using the numerous tools provided by the mathematical theory of chaos. New statistical tests will be used to compare this PRNG to existing ones. Additionally a probabilistic study of its security will be done. Lastly, new applications in computer science will be proposed, especially in the Internet security field.

\bibliographystyle{IEEEtran}
\bibliography{IEEEabrv,Generating_good_chaotic_random_numbers.bib,mabase}

\section*{APPENDIX}

\subsection*{The NIST Statistical Test Suite}

In what follows, the objectives of the fifteen tests contained in the NIST Statistical tests suite are recalled. A more detailed description for those tests can be found in \cite{ANDREW2008}.

\textbf{Frequency (Monobit) Test} is to determine whether the number of ones and zeros in a sequence are approximately the same as would be expected for a truly random sequence.

\textbf{Frequency Test within a Block} is to determine whether the frequency of ones in an M-bits block is approximately M/2, as would be expected under an assumption of randomness (M is the length of each block).

\textbf{Runs Test} is to determine whether the number of runs of ones and zeros of various lengths is as expected for a random sequence. In particular, this test determines whether the oscillation between such zeros and ones is too fast or too slow.

\textbf{Test for the Longest Run of Ones in a Block} is to determine whether the length of the longest run of ones within the tested sequence is consistent with the length of the longest run of ones that would be expected in a random sequence.

\textbf{Binary Matrix Rank Test} is to check for linear dependence among fixed length substrings of the original sequence.

\textbf{Discrete Fourier Transform (Spectral) Test} is to detect periodic features (i.e., repetitive patterns that are near each other) in the tested sequence that would indicate a deviation from the assumption of randomness.

\textbf{Non-overlapping Template Matching Test} is to detect generators that produce too many occurrences of a given non-periodic (aperiodic) pattern.

\textbf{Overlapping Template Matching Test} is the number of occurrences of pre-specified target strings.

\textbf{Maurer's ``Universal Statistical'' Test} is to detect whether or not the sequence can be
significantly compressed without loss of information.

\textbf{Linear Complexity Test} is to determine whether or not the sequence is complex enough to be considered random.

\textbf{Serial Test} is to determine whether the number of occurrences of the $2^{m}$ m-bit (m is the length in bits of each block) overlapping patterns is approximately the same as would be expected for a random sequence.

\textbf{Approximate Entropy Test} is to compare the frequency of overlapping blocks of two consecutive/adjacent lengths (m and m+1) against the expected result for a random sequence (m is the length of each block).

\textbf{Cumulative Sums (Cusum) Test} is to determine whether the cumulative sum of the partial sequences occurring in the tested sequence is too large or too small relative to the expected behavior of that cumulative sum for random sequences.

\textbf{Random Excursions Test} is to determine if the number of visits to a particular state within a cycle deviates from what one would expect for a random sequence.

\textbf{Random Excursions Variant Test} is to detect deviations from the expected number of visits to various states in the random walk.

\end{document}